\documentclass{PoS}

\title{Why not any $\tau$ double bang in IceCube,  yet?}

\ShortTitle{
}

\author{Daniele Fargion\\
        Physics Department, Rome University 1 and INFN Rome1, Ple. A. Moro 2, 00185, Rome, Italy\\
        Technion Institute, Haifa; Israel.
        E-mail: \email{daniele.fargion@roma1.infn.it}}

\author{Pietro Oliva\\
        Department of Electrical Engineering, Niccol\`o Cusano University, Via Don Carlo Gnocchi 3, 00166 Rome, Italy\\
        E-mail: \email{pietro.oliva@unicusano.it}}

\author{Graziano Ucci\\
        Scuola Normale Superiore di Pisa, Piazza dei Cavalieri 7, 56126 Pisa, Italy\\
        E-mail: \email{graziano.ucci@sns.it}}

\abstract{High Energy Neutrino Astronomy has been revealed by a sudden change in the flavor composition above 30 TeV since 2013 by  IceCube detector: the fast growth of spherical showers over atmospheric muon track signal in  IceCube marked the revolution. However these discover didn't led to the promised Neutrino-Astronomy-Land yet.  AGN flaring are not correlated with these high energy tens TeV-PeV events. Brightest persistent or pulsed galactic sources are missing while no point source arise in the lower energy sky. GRB events do not correlate within any minute-hour lapse time windows along any neutrino event. Moreover the astrophysical hard spectra whose exponent was expected at Fermi value of $-2$, seem to converge from $-2.2$ to a softer $-2.7$ or $-3.0$ value, also needed to avoid unobserved  Glashow resonant neutrino at 6.3 PeV energy. Finally a key question arises: why within the ten UHE neutrino, those harder than 200 TeV events (whose timing structure would allow IceCube to  disentangle  any double tau neutrino imprint) don't double bang anyway? We suggest a main solution within a composite flux mostly ruled by prompt atmospheric neutrinos. Nevertheless in the very recent discover of 21 through-going (crossing) muons at hundreds TeVs, whose tracks are more aligned and telling, is shown a first narrow doublet (and some of correlated UHECR clustered source); this points -- or give hints -- for a non-negligible $10-20 \%$ astrophysical component, making neutrino astronomy  already alive anyway.}

\FullConference{Frontier Research in Astrophysics,\\
		26-31 May 2014\\
		Mondello (Palermo), Italy}

\begin{document}
\section{Why High Energy Neutrino Astronomy was hidden for so long?}

Since a century charged Cosmic Rays (CR) are waiting for answers: where and what are the sources and how they get accelerated? Basically, the CR puzzle is based on a deep unsolved riddle: the absence in nature of magnetic monopoles; their eventual existence would fit in the Maxwell equation providing higher symmetry but, unfortunately or fortunately, the magnetic monopole absence is sustained and proved by the large scale magnetic fields existence. The most recent constraint have been defined in IceCube detector by the absence of relativistic (or non-relativistic) bright, heavy monopole radiating tracks.

These magnetic fields that are superimposed onto the spacetime grid in the Galaxy and beyond, are the responsible for the bending of trajectories by charged-currier such as the CR resulting in information loss. Their reconstructed arrival direction is in fact smeared and smoothed in an homogeneous CR sky and their true sources hidden. Actually in last decade among hundred of billions of CR (and gamma) TeVs events, some large scale anisotropy (at a tiny amount $\sim 10^{-4}$ below CR hadrons) have been found in the Milagro, ARGO, HAWC  sky; however, even for us, these traces are relevant and somehow correlated to UHECR \cite{16},  their origination is still controversial and on debate.
 In order to follow the CR arrival directly, we then studied in an indirect way (in the last 50 years) the hard gamma sky, above GeVs, which we assumed to be formed from relativistic electrons via electromagnetic radiation (bremsstrahlung, synchrotron, pair-production, Inverse Compton Scattering -ICS- or self-ICS) but also by hadronic secondaries, namely from $\pi^\pm$, $\pi^0$ decay in ultra-relativistic electron pairs. Indeed, $\pi^0$ decay into neutral $\gamma-\gamma$ channel would lead to a trace,  but also $\pi^\pm$ decay, among the others, would feed neutral neutrinos that would play their game. The different spectra of few sources seem to be in debt to such hadronic component. The Fermi satellite sky at hundred GeV seems to contain both galactic plane quadruple as well as an isotropic probable cosmic component. In addition to these gamma shadows of hadronic components, since 85 years, with the Pauli neutrino proposal and half century ago after neutrino discovery, we hoped to prise out CR secrets through these unique parasite hadronic secondaries, that cannot be born by any other process than nuclear electroweak or hadronic interactions. The astrophysical high energy neutrinos (HE$\nu\gtrsim$ TeVs), once overcame the atmospheric neutrino noise could in fact help us to point back to the CR sources making a very deep and far new Astronomy arise \cite{1}.

 In the last four centuries optics offered, since Galileo, the universe sky map; however since last century, by Maxwell extensions, photon astronomy enriched the view by radio, Infrared, UV, X and Gamma components. The same cosmic big bang noise and the radio or IR relics make the photon sky somehow opaque above TeVs energies, bounding, for instance, PeV photons inside our own galaxy or cutting highest EeV photons at a small fraction (1-2 \%) of  our Universe radius. For more severe reasons, photons cannot escape from the explosive sources at once, but only in late time after a long random walk at the diluted far boundary of the source. On the contrary UHE neutrino may travel promptly from very deep source core and they may travel across the whole universe without almost any opacity.  Low energy (MeVs) neutrinos here on Earth are blazing us because of a huge solar shining flux. At a few tens MeV, among the low energy solar fluxes and the higher energy atmospheric noises, we expect a (still) hidden  narrow cosmic neutrino flux by integrated super-novae neutrinos explosions. A very exciting, but  very probably still blind signals by far SN neutrinos.  However above hundred MeV (possibly up to TeVs)  here on Earth we are polluted by the noisy atmospheric neutrinos. They are secondary traces of the same rich CR rain hitting the atmosphere. They are, as their parent CR, very abundant, but their arrival directions are bent and smeared as CR ones, hiding any lower flux of eventual astrophysical point-sources. The recent homogeneous IceCube neutrino sky at TeVs energy, populated by hundreds of thousands of smeared and featureless signals, is the best testimony of their CR parental origin.

The nearest and brightest neutrinos astronomy recently born since several experiments in a few decades, at MeVs energy, for solar $\nu$s and since 1987 at tens MeV for a rare SN1987a events in LMC, they cannot be extended easily at higher energy because, as we mentioned above, of the CR noise pollution secondaries: bent CR, in fact, poison us with an atmospheric neutrino flux which is dominant over any other astrophysical neutrino signal. Continuous CR flux hitting the terrestrial atmosphere produces, as matter of fact, a persistent smeared and smooth flow of pion, kaons (and muons) whose in-flight decay into neutrinos makes again the neutrino sky a mimic of the blurred and homogeneous CR sky. Nevertheless the highest atmospheric neutrino suffer above TeVs energy of a softer and softer spectra (exponent $-3.7$ up $-4.0$, because of the relativistic survival of muon and pion to their decay). This lower flux  might be overcome by (the expected) harder original astrophysical sources injection spectra, a source spectra as hard as Fermi (exponent $-2$) power law one. Therefore tens TeV up to PeV neutrino astronomy might and indeed is, as it has been discovered, probably opening a new real window to a neutrino sky.
The puzzle is that the consequent sky map and the UHE neutrino  seems still questionable and uncorrelated, offering no astronomy yet.
\begin{figure}[t]
\centering
\includegraphics[scale=0.33]{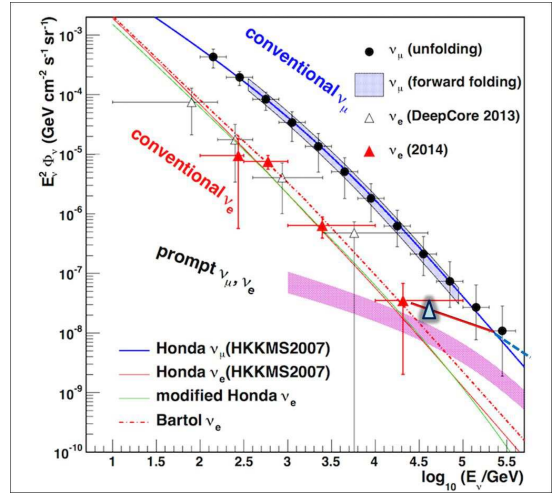}
\caption{The fast neutrino flavor variability of $\nu_\mu$ and $\nu_e+\nu_\tau+\nu_{\mathrm{NC}}$ component at tens TeV energy. The early Deep Core spectra has been recently \cite{5} reinforced by IceCube data making the $\nu_{e}+\nu_\tau+\nu_{\mathrm{NC}}$ showering a very fast appearance, compatible with a hard power law exponent $-2.5$.  A nearly horizontal red line connect recent IceCube spectra by power law exponent $-2.5$. A later dashed green line stand for the   lower astrophysical component at $-2.7$ power. A very similar variability, might be a mimic by a prompt neutrino dominance and a fast composition variability with growing CR energy, around the knee, composition changing from heavy to light nuclei (or nucleon).  This CR behavior is hardening the average nuclear interaction center of mass  energy, leading first to a hard and later to a softer prompt neutrino spectra. The prompt component will feed only or mostly a showering flavor ratio, by $\nu_e$, $\nu_{\mathrm{NC}}$, leading to a $\Phi_{\mathrm{shower}}/\Phi_{\mathrm{track}}$ ratio larger than 2 because of a much larger  $\nu_{e}$ cross section and acceptance in IceCube respect muon ones at those energies. A  prompt neutrino model \cite{11b}  able to combined with such a tiny hardening and a calibrated (the triangle spot) with the observed Icecube events, is shown above. In next figure the corresponding prompt neutrino spectra.   The puzzling of a tau double bang absence  might be  explained by this prompt dominance, where tau are $\frac{1}{20}$ marginal and absent.}
\label{fig:1}
\end{figure}

\section{Remarks on  atmospheric neutrino noise}
\begin{figure}[t]
\centering
\includegraphics[scale=0.4]{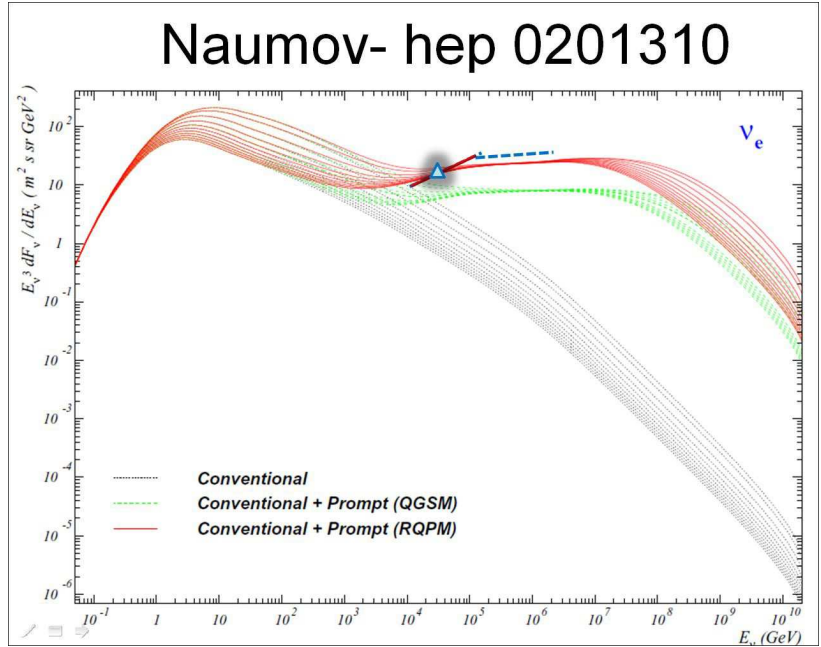}
\caption{As above a schematic draw on a \cite{11b} prompt neutrino model originated by CR  with a late CR composition change from heavy to light nuclei, a change that might fit, as shown in figure the IceCube hard to soft spectra.
Indeed a power law exponent $-2.5$, might be offered (in a narrow windows) by a prompt neutrino dominance with a composition variability with growing CR energy around the knee, composition changing from heavy to light nuclei (or nucleon). This CR behavior is hardening for a while, the average nuclear interaction energy, tuning a hard  $\sim-2.5$ spectra from tens-hundreds TeV energies, followed by a softer  power law exponent $-2.67$,$-3$ as the CR one, at higher energies. The new prompt components will feed only or mostly a showering flavor ratio due to $\nu_{e}+\nu_{\mathrm{NC}}$, leading to a $\Phi_{\mathrm{shower}}/\Phi_{\mathrm{track}}$ ratio  larger than 2 because of much larger $\nu_{e}$ cross section and acceptance in IceCube. Tau signals as well as their double bang above 200 TeV energy, might be avoided}
\label{fig:2}
\end{figure}

\begin{figure}[t]
\centering
\includegraphics[scale=0.45]{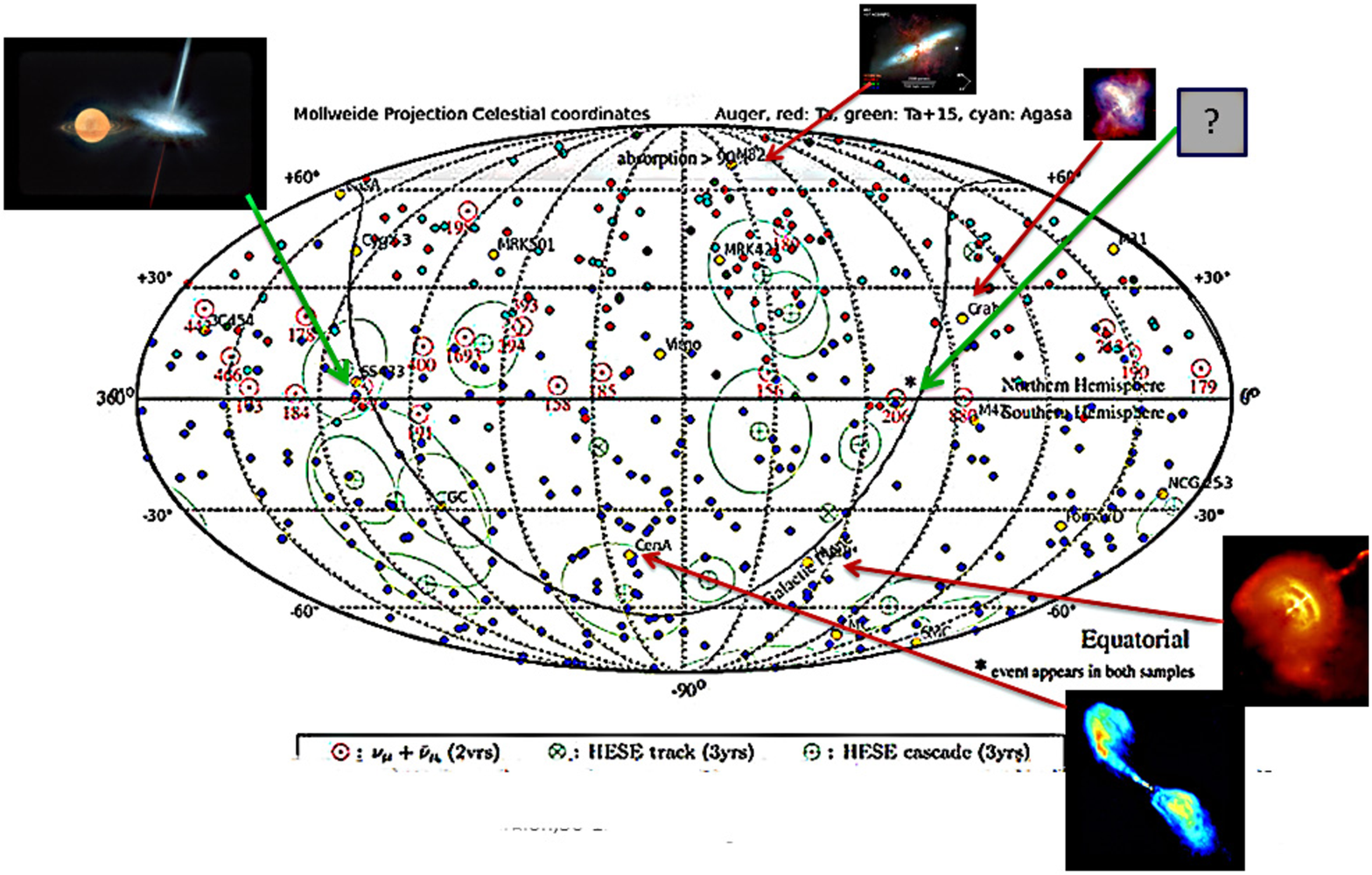}
\caption{UHECR by AUGER (blue dots), TA (red dots) and AGASA (magenta dots) versus new IceCube events: the latter are label with $\oplus$ for showers (16), $\otimes$  for contained HESE tracks ( just $3$), $\odot$ for trough going muons ($21$). In the figure are shown some of the main expected UHECR correlations; the red arrow point to absence of correlations; the green ones point to first candidate sources. Note in particular the SS433 binary jet and the n.5 IceCube event near galactic plane marked by a small asterisk. These two correlation are promising a non negligible neutrino astronomy within a (unfortunate, just discovered) prompt neutrino noise.}
\label{fig:3}
\end{figure}

If the Neutrino astronomy did not correlate (yet) to gamma photon sky, how did  IceCube recognized and claimed anyway an astrophysical signals? The sudden neutrino flavor ratio evolution offered the key (see Fig.\ref{fig:1}).
As we have noted the atmospheric neutrinos completely sink any astrophysical information in a sea of noise. There is however a very characteristic imprint of $\pi^\pm$ and $\mu^\pm$: the lifetime of $\pi^\pm$ (and of K$^\pm$) is short, of the order of magnitude of $10^{-8}$ seconds, and its decay favorites the $\mu^\pm$ channel instead of the electronic one (because of helicity arguments) which then favor the presence of $\nu_\mu$, $\overline{\nu}_\mu$ over  $\nu_e$, $\overline{\nu}_e$. The muons instead are longer living compared to pions: $\mu^\pm$ mean lifetime is a hundred of time longer than that of $\pi^\pm$ ($t^{\mu^\pm}_{{}^{1}\!/_{2}}\sim2\cdot10^{-6}$ s while $t^{\pi^\pm}_{{}^{1}\!/_{2}}\sim2,6\cdot10^{-8}$ s) and their decay leads (up to a threshold near tens GeV) to twice as much of neutrinos ($\mu^+\to e^++\nu_e+\overline{\nu}_\mu$ and $\mu^-\to e^-+\overline{\nu}_e+\nu_\mu$). This imply that at few GeV energies the ratio $\phi_{\nu_\mu}:\phi_{\nu_e}$ is expected to be around 2, neglecting for a moment for the oscillation across the Earth. Oscillations across the Universe that would be the key of the  IceCube flavor revolution.
This ratio 2 among muon and electron has been well verified for vertical neutrino fluxes. However for upgoing neutrinos, coming from below through the Earth, the mentioned ratio has proven to be surprisingly close to 1, this being then the main imprint for the $\nu_\mu\to\nu_\tau$ averaged mixing confirming what it has been somehow foreseen by Pontecorvo in 1957 about the eventual $\nu_\mu\to\nu_e$ flavor mixing.
We can now, for a moment, neglect this $\nu_\mu\to\nu_\tau$ mixing (that become less and less probable crossing the Earth size, as energy rises well above tens GeV); the non-decay of energetic muons makes the atmospheric $\phi_{\nu_\mu}$ survival, by pion decays,  more and more abundant over $\phi_{\nu_e}$, just because of the premature  suppression of $\mu$ decay in flight above a few ($\sim4$ GeV) energy range. Indeed at these energies the muons travels too long ($\sim20$ km) producing no more electronic neutrinos, while pions and kaons are still decaying into muons and the same flavor neutrinos $\nu_\mu$. This mechanism makes $\phi_{\nu_\mu}$ larger and larger ($\propto$ E) over the $\phi_{\nu_e}$. Nevertheless the contribute of inclined events may still feed a little the $\phi_{\nu_e}$ flux,(see Fig.\ref{fig:1})(see Fig.\ref{fig:2}).

Above a few hundred GeV pions themselves (along with harder kaons $K^\pm$) do not have time to decay in air, consequently $\phi_{\nu_\mu}$ suffers also of a similar (as electron neutrino) suppression, asymptotically freezing to a $\phi_{\nu_\mu}:\phi_{\nu_e}$ ratio around 20. This theoretical prediction has been confirmed by DeepCore (inner core of IceCube detector able to track tens GeV signals) and in IceCube itself  in the last year (2014), where from 100 GeV up to 10 TeV the $\phi_{\nu_\mu}:\phi_{\nu_e}$ ratio showed to be just 16$\div$20. The muon neutrino leaves a long track in  IceCube, while the electron makes an electromagnetic shower tree, of several meter long in ice,  whose final  outcome inside a (diffuser, but transparent) cubic km detector is a spherical (hundreds meter size ball) shower, whose lightening is mostly made by randomized Cherenkov photons. Just above this energy was found the IceCube revolution that made the last three years of IceCube the most exciting in neutrino detector on Earth: the blow of showers over muon tracks.
However above hundred TeVs also charmed mesons may arise and their decay in fast timescale may feed the additional prompt atmospheric neutrino \cite{10}, \cite{11} whose flux might mimic a CR one because their rapid decay into prompt neutrino flavors ($\nu_{e}$, $\nu_{\mu}$, $\nu_{\tau}=1$,~1,~$\simeq \frac{1}{20}$) \cite{9}. Therefore prompt neutrino dominance may explain present, puzzling double bang tau absence \cite{0}, as it is somehow described by figures above, based on early prompt neutrino expectations assuming a little hardening in the prompt spectra about 30 TeV \cite{11b}.

\section{Three-flavor mixing in Universe feeding astrophysical $\nu_\tau$ }

The Kamiokande $\nu_\mu\to\nu_\tau$ discover (1997-98) has been the first main lepton flavor mix proved and it has been independently reinforced by SNO and other (like earliest GALLEX) results on solar neutrino fluxes. These results are the most recent reasons for the 2015 Nobel prize based on the amazing neutrino mass and its mixing summarized by the Pontecorvo-Maki-Nakagawa-Sakata matrix. The power of these neutrino discoveries are the hint for an important $\nu_\tau$ component very useful to Astrophysics and the main argument of our article. Indeed, even within the present negligible neutrino mass splitting ($\sim\Delta\mathrm{m}_{\mathbf{atm}}\simeq5\cdot10^{-2}$~eV and $\Delta \mathrm{m}_{\mathbf{solar}}\simeq8.5\cdot10^{-3}$~eV) and within the high energies considered in astrophysics (TeV, PeV, EeV), the inter stellar and the galactic distances are large enough to guarantee a complete averaged flavor mixing even if the starting $\nu$s are mostly from $\pi$ decays, so that in spite of the initial $(\phi_{\nu_e}:\phi_{\nu_\mu}:\phi_{\nu_\tau})=(1:2:0)$ ratio, we will find a final ratio of (almost) democratic distribution $(\phi_{\nu_e}:\phi_{\nu_\mu}:\phi_{\nu_\tau})=(1:1:1)$ \cite{17} \cite{6}. This expected flux in flavor is, possibly, the main reason for the more recent revolution in IceCube since November 2012: the PeV (or later found hundred TeVs) neutrino showering versus corresponding muon tracks are 4 times more abundant than the poor ratio 1/20 at TeV energies.
Indeed the three PeV events in last four years and the additional few tens (or hundred) TeV $\nu$ in IceCube made an exceptional discover: the $\phi_{\nu_\mu}:\phi_{\nu_e}$ ratio originally at $\sim20$ (at 10 TeV) must change almost at once,   toward the value of 1/3 or 1/4 and this change must happen quickly suggesting a very hard spectra. So, the first 27 $\nu$ events (2013) and later or present 54 (2015) events are mostly showers (39 shower  produced by an hadronic or electromagnetic cascade in ice) and only nearly a dozen ($15-2= 13$)  of events are $\nu_\mu\to\mu$ tracks. These are the main claim of a novel $\nu$ signal and extraterrestrial (possibly) astronomy. It should noted that a more recent (6 November, 30 November 2015)  IceCube articles \cite{2} excluded 6 more HESE (High Energy Starting Events) from the list (from 15 just to 7 HESE)  while they are including 9 through going muons in their maps; the total number 55 remind the earlier 54 but it is drastically different.
These new IceCube maps have been overlapped on UHECR maps of AUGER and TA looking for correlations.
As we anticipated, last ICRC articles rejected any connection while we did found (and reconfirm) several correlated sources (see Fig.\ref{fig:3}).

Therefore late 2012 and recent 2013-14-15 IceCube signal have been the running dates of this new high energy $\nu$ astronomy leading to a huge list of (often) non-converging models of the HESE neutrino events trying to match the correlation with the families of sources: AGN or BL Lac  \cite{7}, Dark GRB, Galactic and Cosmic sources as well as cosmic relic heavy decay particle \cite{8}. However we faced these unanswered questions that we must give answers to:
\begin{enumerate}
\item Why, up to now, there is no clear signature of our galactic plane? Most Fermi and TeV $\gamma$~telescope astronomy do show anyway both extragalactic but also a clear galactic plane signal. One may mention a PeV shower event centered at Galactic Center direction but there are not too many galactic signals within last 54 events (just as analogy: Gamma Ray Bursts -GRB- are isotropic but their twin Soft Gamma Repeaters are galactic and are not negligible).
\item  Why there is no sharp self-clustering among the highest 54 events (likewise in the TeV IceCube maps)?
\item  Why there is no any sharp repeater (out of the newest doublet discussed below)?
\item  Why no GRB-$\nu$ time connection arises?
\item  Why there is no point-source or known $\gamma$ source candidates (except the SS433 discussed below) connections with $\nu$s (limits from ANTARES and IceCube) \cite{3}?
\item  Why there is no expected (and preliminary claimed) Fermi power spectrum ($\gamma=-2$) but a softer power law, pointing at the more familiar $\gamma=-2.67$, a suspiciously mimic of the CR spectrum?
\item Why we don't see the Glashow resonance at 6.3 PeV needing for a knee in UHE neutrino spectra?
\item Why there's no correlation with persistent AGN flaring $\gamma$ TeV sources \cite{7}?
\item In particular, within all the 10 events at energy above two hundreds-TeV IceCube $\nu$ that might be disentangled in their time structure, why among the them there is no  $\tau$ signal \cite{0}? Let us remind that, on the contrary,  there have been observed three muon HESE neutrino whose detection area in these energies is ten times smaller than tau one \cite{0}.
\end{enumerate}

\section{The UHECR versus UHE-neutrino maps}

There are good reasons to believe that an UHECR  may become soon an astronomy  \cite{13}: GZK cut off make them contained in a narrow Universe; UHECR rigidity offers somehow directionality for the source. Since two decades this hope is becoming a reality by the high AUGER-TA statistics. However in other more recent studies the correlations (as the North or South  HOT SPOTS)  has not been confirmed \cite{2}. The $55$ EeV UHECR (as well their EeV traces) being fragmented and decayed in flight  may also shine secondaries in tens EeV energy range, as it has been foreseen and later observed around Cen-A twin train of events \cite{17b}; the same fragments being UHECR and possibly radioactive they might decay and shine in hundred TeV-PeVs energy too, making the observed TeVs anisotropy explained  \cite{16}. Therefore there is a good reason to try to overlap UHECR and neutrino signals, both the  showering but mainly the muon tracks.

Indeed these tracks could be better connecting to the sources \cite{12} than any wide spread shower ($15^\circ$) because of their much smaller resolution trajectory ($1^\circ- 0.4^\circ$) at hundreds TeVs. Few tens of these events may be testing a neutrino astronomy birth.
In particular recent IceCube HESE (High Energy Starting Events) $7$, and muon crossing $9$ and showering events $39$, ($7+9+39= 55$) has been published in \cite{4}, \cite{2}; however in the same IceCube presentation on the web site,  there have been shown many more through going events in a higher energy selection, specifically $21$, $3$, $16$, ($21+3+16= 40$). Namely, the map shown above had shown the degree of overlapping of the UHECR clustering noting \cite{13} a number of UHECR triplet in a narrow area. In particular the SS433 source has been pointed out since a long time, while the earlier HESE event n.5 in IceCube map had been found to overlap in the same map by a new through going event (see Fig.\ref{fig:3}).

\subsection{Probability }
It is easy to show that the tau absence is quite un-probable (even within ten observed events) at a level of $4.4\%$ as well as it is quite unrealistic that $3$ muon tracks have been observed in tau absence. That make the tau double bang problem real. As it has been foreseen the narrow muon track doublet (see Fig.\ref{fig:3}) has a very interesting probability to occur \cite{12}. There are two ways to make a correlation: a priori the twin narrow self clustering within 24 events has a probability $P$ to not occur as large as \cite{12}
$$
P = e^{-\varepsilon\cdot \frac{n\cdot(n-1)}{2}}
$$
where $\epsilon$ is the ratio among all the allowable neutrino sky (because of the Earth opacity and the half up-going sky, just $33\%$ of the whole $4\pi$) and the tiny track resolution $0.5^\circ$. This leads because of $\epsilon\simeq6\cdot 10^{-5}$, $n= 24$,  to a probability to not find any of such doublet of $P=98.35\%$, that is to say the probability to find such a doublet is below $1.6\%$.

The probability to find a crossing muon at the center view of SS433 $\pm3^\circ$ source, that we identified as a candidate source  by clustering of most powerful  UHECR events (triplet), is near or below  $1\%$.
Therefore a neutrino clustering might correlate with UHECR.

\section{Conclusions}

 About the randomness of the present understanding of IceCube events we may address the reader to several dozens (or hundreds) of models and articles trying to find a successful fit candidate for the $\nu$ events, each one of them often going in opposite direction with respect to others: from cosmic relic decay of some new particles \cite{8} to hidden or opaque AGN or GRB, models mainly galactic or cosmic-only scenarios. A great market of proposals and solution that are necessarily frequently in plain conflict with each other.
One of the IceCube puzzle (or conflict) is the impossibility to extrapolate tens TeV-PeV IceCube $\nu$ spectra with the Fermi cosmic $\gamma$ background. Within 54  $\nu$ events contained in IceCube, the dozen of $\nu_\mu$ high energy track ones, most require several dozen of similar crossing $\mu$ generated outside the km$^3$ detector; the tens TeV-PeV are several km ($\gtrsim10$ km) long, amplifying their direction. The rare doublet is a promising first milestone in astronomy road.

\subsection{The charm atmospheric contribute versus astrophysical one}

The main solution we foresee at the moment, with all the reservations, is the following one:
most of the HESE signals are still noisy atmospheric neutrino but of charmed nature;
they are the prompt neutrinos. This solution seems not widely accepted. Most prompt model
have foreseen lower (even factor five) below, but most models admit an order of magnitude in error bars. These model mostly require a prompt charmed power spectrum exponent of $-2.67$ that mimics our cosmic ray one.
Therefore the presence of diffused random signals with no correlation may be well understood.
The flavor component from common muon dominated  to a charmed ratio $(\nu_e:\nu_{\mu}) = (1:1)$
may explain (with the help of Neutral Current component) the composite ratio Track:Shower = (1:1+1).
This ratio may be amplified to Track:Shower = (1:4) by the better detection area of electron and its higher energy release. Finally the absence of tau neutrino is guaranteed by the negligible production of
tau flavor in tens TeV- PeV energy range as prompt neutrino $(\nu_e:\nu_{\mu}:\nu_{\tau})=(1:1:\frac{1}{20})$.

Naturally most of the scientific model babel may be understood. The apparent disagreement of the early $-2.2$ power exponent versus $\-2.67$ at higher energy could be understood assuming at PeVs-tens PeV (the knee) a CR composition transition from heavy nuclei toward light ones. That composition transition introduces a fast average nucleon energy change that is reflected into a hardening of the spectra within the narrow range of $10-100$ TeV.

\subsection{Foreseen tau signals, anyways}

In a brief summary, the Icecube didn't find double tau bangs because they discovered mostly prompt neutrino signals where only  $\nu_{\mu}$ and  $\nu_{e}$ are produced. There is anyway almost 10\%, 20\% of present  IceCube signal, at ${\Phi_{\nu}\sim1\leftrightarrows\,2\;\mathrm{eV}\cdot\mathrm{cm}^{-2}\cdot \mathrm{s}^{-1}\cdot \mathrm{sr}^{-1}}$, a very reasonable contribute of astrophysical neutrinos, that finally may arise.
This flux is coherent with the Fermi extrapolated diffused gamma sky above TeV. Therefore some correlation with expected sources may soon rise.
Indeed a first UHE $\nu_{\mu}$ neutrino doublet (within a very narrow solid angle) made by HESE track (event n. 5, Fig. \ref{fig:3}) is overlapped with a through going muon  event at 200 TeV; it  has been found and shown recently  (but not published yet) in IceCube web site. The probability to occur by chance within 25 neutrino track events is quite small ($\lesssim$1\%), therefore at least a small fraction 10\%, 20\% of the events might be of astrophysical nature, but, for tau absence,  most are prompt signals. Within a few dozens of two hundreds TeVs events a tau double bangs may (and it must anyway) rise in next few years. But the flavor ratio, if we are right, will sound (as already it does) extremely improbable for any dominant ``democratic in flavor'' astrophysical model. The  astrophysical tiny 10\%, 20\% presence, a possible source of first UHE neutrino doublets, gives hope for the first real birth of the Neutrino Astronomy, unfortunately or fortunately, below the new, additional noisy ashes of the present charmed  prompt atmospheric neutrino. The need of a more filtered neutrino astronomy, after all, pushes us to remind that our proposal of a Tau air-shower astronomy is more than ever actual for a noise-free astrophysical neutrino astronomy that  may occur at PeVs energies and above by UHE neutrino scattering inside a mountain (or upward in  Earth crust \cite{21}) and by its tau, crossing hundreds meters, exiting from the rock and decaying in flight, in air \cite{22}. This $\tau$ decay leads to a Tau-airshower spreads (horizontal or upward) into more than a billion photon signals whose Cherenkov \cite{19} (or radio, as new proposed GRAND experiment) photons (as well as its million electron-pairs component)  are the best amplified and widely spread signals in large area, noise-free signature of a tau neutrino astronomy. These tau airshower detection project (as AUGER, TA, ASHRA, GRAND)  should  be more supported and developed \cite{19}, \cite{20}, \cite{18}, \cite{8a}.

\section{Acknowledgments}
The authors gratefully acknowledge the referee Prof. T. Stanev for his valuable comments which helped to improve the manuscript.

\end{document}